\documentclass[pra,a4paper,twocolumn,showpacs]{revtex4}
\usepackage{graphicx}
\usepackage[latin1]{inputenc}

\newcommand{\K}[1]{$^{#1}$K}
\newcommand{\Rb}{$^{87}$Rb}
\newcommand{\ket}[1]{| #1 \rangle}

\begin{document}

\title{Collisional properties of sympathetically cooled \K{39}}

\author{L. De Sarlo\footnote{Corresponding author: desarlo@lens.unifi.it}, 
P. Maioli\footnote{Present address: Laboratoire de Spectrométrie Ionique et Moléculaire (LASIM), Université Claude Bernard Lyon 1, France }, G. Barontini, J. Catani$^1$,
F. Minardi$^{1,2}$, and M. Inguscio$^{1,2}$}

\affiliation{LENS - European Laboratory for Non-Linear Spectroscopy and Dipartimento di Fisica, Università di Firenze,
via N. Carrara 1, I-50019 Sesto Fiorentino - Firenze, Italy\\
$^1$INFN, via G. Sansone
1, I-50019 Sesto Fiorentino - Firenze, Italy\\
$^2$CNR-INFM, via G. Sansone 1, I-50019 Sesto Fiorentino - Firenze, Italy}

\date{\today}

\begin{abstract}
We report the experimental evidence of the sympathetic cooling of \K{39} with \Rb\ down 
to $1 \, \mu$K, obtained in a novel tight confining magnetic trap. This allowed us to
perform the first direct measurement of the elastic cross section of \K{39} below 
$50 \,\mu$K. The result obtained for the triplet scattering length, $a_T = -51(7)$ 
Bohr radii, agrees with previous results derived from photoassociation spectra and from 
Feshbach spectroscopy of \K{40}.
\end{abstract}
\maketitle
\section{Introduction}
In the field of ultracold and quantum gases Potassium gained a key role since it is the
only alkali, beside Lithium, for which a stable fermionic isotope exists. 
Furthermore, since the sympathetic cooling of fermionic \K{40} with \Rb, the most 
widespread atom in the laboratories devoted to laser cooling, is particularly favorable, 
it is not surprising that so many experiments with ultracold fermions were carried 
out or are planned with this atom. In contrast very little 
attention was paid to the two other (bosonic) isotopes of Potassium,
$^{39}$K and $^{41}$K. These isotopes offer the possibility of creating double 
species Bose-Einstein condensates, which display rich
and interesting quantum phase diagrams when trapped in optical lattices \cite{AltmanSpinBECPD,Isacsson2BosonsOL,WuQPT,ZhengPhaseDiagTBEC}. 
After the seminal work where a double 
species $^{41}$K-\Rb\ BEC was produced in a magnetic trap \cite{ModugnoTBEC}, 
essentially no experiments have addressed these topics. 

Along these guidelines, we started an experiment devoted to the exploration of degenerate
Bose-Bose mixtures in optical lattices. In this work we report that also $^{39}$K can be 
sympathetically cooled with \Rb\ although this is much more difficult due to its 
unfavorable laser cooling and scattering properties, thus demonstrating that all the 
stable isotopes of Potassium can be cooled to ultralow temperatures. 

Laser cooling of \K{39} has several disadvantages if compared to alkalis such as \Rb\ or
even \K{40}. These disadvantages are related to the hyperfine 
structure of the $P_{3/2}$ level whose separation is of the same order of magnitude as 
the natural linewidth of the cooling transition 
$\ket{F=2,m_F=+2} \rightarrow \ket{F=3,m_F=+3}$. The resulting strong optical pumping towards
the $\ket{F=1}$ manifold must be countered by a repumping light with an intensity 
comparable to that of the light driving the cooling transition. A careful balance of the
intensity and the detuning of these two laser frequencies, unsuitable for optimized 
loading of a Magneto-Optical-Trap (MOT), is the only way to obtain temperatures at the 
Doppler limit by laser cooling \cite{FortBambini}. 
Furthermore, the s-wave elastic cross section of \K{39} at zero temperature is more than 
one order of magnitude smaller than that of \Rb\ and, due to the attractive character of
the interaction, the Ramsauer-Townsend minimum \cite{MottMassey} occurs at a temperature 
of around $320 \,\mu$K where the contribution of other partial waves is still small
\cite{simoni}.
All these features strongly hamper the efficiency of
evaporative cooling. Note that, due to Pauli blocking, an even more
dramatic decrease in the evaporative cooling efficiency occurs for the fermionic \K{40}.

For these reasons the most widely employed experimental technique to reach ultralow 
temperatures in a K gas is sympathetic cooling with \Rb. This amounts to producing 
a mixed MOT of the two species and then evaporatively cool \Rb\ allowing K to 
thermalize with it. With this technique it has been possible to reach quantum 
degeneracy of \K{40} \cite{JinFermiGas} and \K{41} \cite{ModugnoKBEC}, but no attempts
have been made to test this method on \K{39}. For this mixture in fact, the inter-species
cross section is again more than one order of magnitude smaller than the one of 
\Rb-\K{40} and \Rb-\K{41} \cite{FerlainoFeshbachKRb,FerlainoFeshbachKRbERR}.

We demonstrate here for the first time that \K{39} can be sympathetically cooled 
to $1 \,\mu$K. 
Therefore the work presented in this paper is a critical step towards 
the production of a Bose-Einstein Condensate (BEC) of \K{39}. Such a system is a
promising candidate for the realization of a BEC with interaction tunable around zero,
since a broad Feshbach resonance is predicted around $400\,$G
\cite{FerlainoFeshbachKRb,FerlainoFeshbachKRbERR}. 

The achievement of sympathetic cooling allowed us to perform the first direct 
measurement of the s-wave elastic cross section of \K{39}. Assuming from earlier work
on photoassociation spectra \cite{Gould39K} and on Feshbach spectroscopy of \K{40}
 \cite{JinFeshbachK40} the attractive character of the interatomic interaction 
(i.e. the negative sign of the s-wave scattering length), our results agrees with 
\cite{Gould39K,JinFeshbachK40} within $1.7$ combined standard deviations.

The time required for sympathetic cooling depends crucially on the collision rate
between the two species. If this time is of the same order of magnitude as the
lifetime of the sample, the density will drop during the evaporation and cooling 
will eventually stop. 
Since the cross section for \K{39}-\Rb\ collision is smaller than in the case of \K{40}
and \K{41}, one has to increase the lifetime of the sample and/or the confinement
before starting sympathetic cooling.  
In the experiments reported here, the confinement of the \K{39} sample is
increased by more than a factor $2$ with respect to \cite{ModugnoTBEC} employing 
a new kind of magneto-static trap of size intermediate between micro-traps, where 
microscopic current-carrying wires are laid on a chip, and traditional magnetic traps
created with multiple winding coils \cite{MinardiMTrapStanford}.

The outline of the paper is the following: in section \ref{sec:exp} we will present 
a brief description of the experimental apparatus and the results about sympathetic 
cooling of \Rb\ and \K{39}. In section \ref{sec:meas} we will introduce the
experimental procedure for studying ultracold collision in \K{39} and  we will present 
the theoretical analysis of the data.
Finally, we will conclude with the perspectives of \Rb - \K{39} evaporative cooling
and of a \K{39} BEC.

\section{\label{sec:exp}Experimental Apparatus}
\subsection{Laser system and MOT loading}
In order to provide the maximum flexibility for the loading of the mixed MOT, our
experimental apparatus features two distinct two-dimensional MOT's (2D-MOT) for the
separate precooling of \Rb\ and \K{39}. A complete description of our laser system 
and a thorough characterization of our \K{39} 2D-MOT can be found in 
Ref. \cite{CataniK2DMOT}. The 2D-MOT employed for \Rb\ has exactly the same structure and 
similar performances. In the experiments we used $130\,(110)$ mW for the
cooling light and $40\,(6)$ mW of repumping light for \K{39} (\Rb). The 
atomic flux provided by the two 2D-MOT's is collected inside a mixed MOT formed in a 
ultra high vacuum chamber. 
The MOT is formed by a standard configuration with six independent laser beams with 
$25$ mm diameter and a total power of $45\,(40)$ mW for the cooling light and 
$24\,(6)$ mW for the repumping light of \K{39} (\Rb). These beams are split from 
the output of one single mode fiber in which we inject the four different
frequencies needed for the two species, thereby guaranteeing a perfect geometric overlap
of the cooling and repumping light and of the two MOT's. The magnetic field gradient of
about $16\,$G/cm used in the MOT is provided by a pair of anti-Helmholtz coils operating
at $4\,$A. 

We typically load around $2 \times 10^9$ atoms of \Rb\ and $5 \times 10^6$ atoms of
\K{39} in $10$ and $1$ s respectively. We can control very precisely the number of
\K{39} atoms loaded in the MOT adjusting the time of operation of the 2D-MOT. 
After the loading of the MOT we reduce the repumping intensity and increase the 
magnetic field gradient to compress the cloud, we apply a $3.5\,$ms stage of 
polarization-gradient cooling, optically pump the atoms of both species towards the 
$\ket{F=2,m_F=+2}$ state and then trap them in a purely magnetic trap. 

\subsection{Magnetic trap and evaporation}
Our magnetic trap represents a compromise between the usual magnetic traps generated by
coils placed outside the vacuum system, typically low confining ($\nu_{\mathrm{max}}
\sim 100\,$Hz), high power consuming ($\sim 1\,$kW), several centimeters in size, which 
produce the largest BEC's ($> 10^6$ atoms), and the so called microtraps 
operating in vacuum, providing a high confinement ($\nu_{\mathrm{max}} \sim 1$ kHz),
requiring low electrical power ($\sim 1\,$W), that easily fit on a microchip, but usually
produce smaller condensates ($\sim 10^4$ atoms). 

The potential generated by our trap is of the Ioffe-Pritchard type and the 
structure has both current conductors laying on a chip and free standing 
\cite{MinardiMTrapStanford}. The free
standing conductors are built from a single oxygen-free copper tube machined to 
form the four Ioffe bars and two partial rings that provide the end caps of the
potential in the axial direction. The axial confinement is increased by a circular 
copper trace laid on a direct-bond copper chip that interfaces the current 
leads and the free standing conductors. Electrical and mechanical contact between
the parts is obtained by vacuum brazing and the current leads are brought outside
the Ultra-High Vacuum environment by a custom-made electrical feedthrough. To allow 
optical access along the horizontal axial direction, the feedthrough is hollow and 
terminates with a vacuum viewport; optical access in the radial direction is easily 
obtained through the $2$ mm gaps between the Ioffe bars. Since these gaps are too 
small to allow the efficient production of a MOT, the magnetic trap, hereafter 
called millimetric-trap or \emph{mTrap}, is displaced by $26$ mm from the center of 
the MOT. A big coil placed outside vacuum is used to adjust the bias field of the trap 
thereby changing the radial confinement. The confinement provided by our mTrap at $95$ A 
current and $6$ G bias field is $\omega_r =2 \pi \times 447(7)$ Hz and 
$\omega_z = 2 \pi \times 29.20(1)$Hz for \K{39}.

The displacement of the atoms from the MOT position to the mTrap is accomplished holding 
them inside a quadrupole trap which is moved with
a motorized translation stage. The quadrupole trap is formed ramping the current of the 
MOT coils up to $65$ A and the motion takes less than $500$ ms in order to reduce the 
losses due to Majorana spin-flip. 

While the magnetic field of the mTrap adds to that of the transfer trap along 
the radial directions, along the axial direction the mTrap generates a field 
directed in the opposite direction so that an adiabatic transfer of atoms would 
be only possible decreasing the trap depth during the process and therefore 
reducing the number of transferred atoms. To overcome this potential pitfall we 
employ a mixed approach to load the trap: we ramp up the current of the mTrap 
in $150\,$ms, a time which is fully adiabatic only for the radial direction. 
The overall efficiency of the transfer from the MOT to the mTrap is around 30\%, 
and we typically trap $7 \times 10^8$ \Rb\ and $4 \times 10^5$ \K{39} atoms. 

\begin{figure}
\includegraphics[width=220pt]{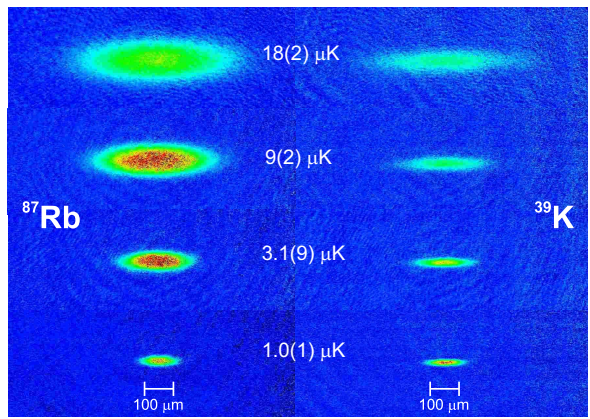}
\caption{(Color online) Experimental sequence showing sympathetic cooling between 
\Rb\ (\emph{left}) and \K{39} (\emph{right}). All the images are taken after 
$1$ ms expansion. The phase space density at $1\, \mu$K is $0.02$ and $0.01$. 
Images are taken along the vertical radial direction.
}\label{fig:symp}
\end{figure}

Once atoms are loaded inside the magnetic trap we perform evaporation on the \Rb\
cloud by transferring the hottest atoms to the antitrapped $\ket{F=1,m_F=+1}$ state
with a microwave sweep around 6.8 GHz. In addition we apply a second ramp to remove
unwanted atoms from the $\ket{F=2,m_F=+1}$ level. The \K{39} is unaffected by these
sweeps, owing to the huge difference in frequency with the corresponding hyperfine
transition (460 MHz), and only minor losses occur during the evaporation. Provided that 
the number of \Rb\ atoms $N_{Rb}$ at a given temperature is still larger than the 
initial number of \K{39} atoms $N_K$, the former can act as coolant establishing thermal
equilibrium, as it is shown in Fig.~\ref{fig:symp}. For this reason the initial
number of \K{39} atoms is reduced in the pictures from $2.7 \times 10^5$ in the first 
picture to $5.9 \times 10^4$ of the last one at a temperature of $1.0(1)\,\mu$K.  

We experimentally found that a sufficient condition to reach thermal equilibrium is 
$N_{Rb} / N_K  > 2.7$, but no attempt has been made to measure the inter-species elastic 
cross section since it can be obtained with high precision from the results
of Feshbach spectroscopy of the 
\K{40}-\Rb\ mixture \cite{FerlainoFeshbachKRb,FerlainoFeshbachKRbERR}. We used instead the 
known value $a_{K-Rb} = +36 a_0$ to guide the optimization of the evaporation.
Indeed we experimentally verified that in presence of \K{39} atoms, in order to 
optimize the number of \Rb\ atoms below a few $\mu$K, the evaporation ramp must be 
slowed down in the last part to be sure that the two species are always in thermal 
equilibrium.

\section{Cross-thermalization measurement}\label{sec:meas}
\subsection{Introduction}
\begin{figure}
\includegraphics[width=220pt]{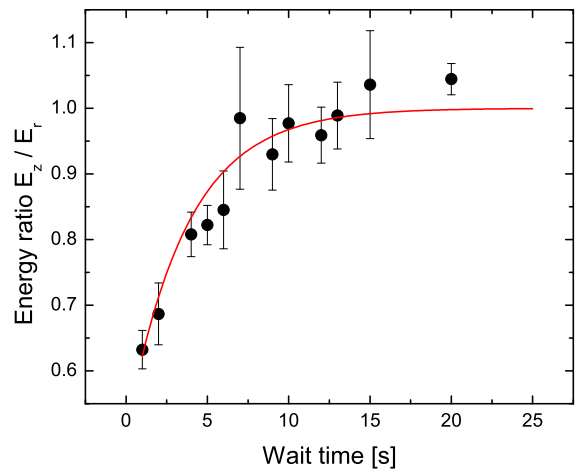}
\caption{(Color online) Plot of the relaxation dynamics of an ultracold sample of 
\K{39} after parametric heating in the radial direction. Data are taken after expansion 
(see text). Each point is an average of several experimental realization and the
line is an exponential decay fit with equilibrium value fixed to $1$, 
initial value and time constant as free parameters. Average initial number of atoms is 
$379(10) \times 10^3$ and initial average temperature is 
$16.2(7)\,\mu$K. From the fit we obtain $\tau = 3.66(46)\,$s.}\label{fig:relax}
\end{figure}

We measure the cross section of \K{39} by analyzing the relaxation of the atomic cloud towards thermal equilibrium. The technique can be described as follows.
The dynamics of an ideal gas isolated and confined in an harmonic potential is 
completely separable meaning that the different spatial degrees of freedom are decoupled. 
If such a sample is prepared with different energy distributions along different 
directions this difference will not vary with time. On the other hand, due to 
interactions, the collisions drive the system towards thermal equilibrium, 
namely a state in which the total energy is conserved and the effective 
temperatures of the different degrees of freedom are equal. If the initial 
state has different effective temperatures, the relaxation towards thermal equilibrium 
can provide direct information on the collisions. In our system this is 
accomplished in the following way: after
obtaining a cloud of cold \K{39} as described in Section \ref{sec:exp}, \Rb\ is blowed
away from the trap with a pulse of resonant light and the radial degrees of freedom 
of K are excited by parametric excitation, obtained modulating the value of 
the trap bias field at the frequency of radial confinement for $100$ ms. 
Due to our elongated trap geometry this excitation is highly  
selective and the axial degree of freedom maintains its effective temperature. 
Following this excitation we allow the cloud to relax for a time $t_w$ after 
which we image the cloud along the vertical radial direction by means of 
\emph{in situ} absorption imaging. From the measured widths $w$ of the Gaussian 
density profile we can calculate the average potential energy along the radial ($r$)
and axial ($z$) direction as
\begin{equation}
E_i = \frac{1}{2} k_B T_i = \frac{m}{2} (\omega_i w_i)^2 \quad i=r,z \,\, .
\end{equation}
where $m$ is the atomic mass. 
Repeating this measurement for different $t_w$ allows us to measure the relaxation of the
ratio between the axial and radial temperatures. As shown in Fig.~\ref{fig:relax} this is
well fitted by an exponential decay. The time constant $\tau$ of this decay is related to
the elastic scattering cross section and to the density of the sample by the following
relation:
\begin{equation}\label{eq:gscatp}
\tau^{-1} = \frac{\gamma_{el}}{\alpha} = \frac{\bar n \langle \sigma v 
\rangle}{\alpha} ,
\end{equation}
where $\alpha$ is the number of collisions required to an atom to reach thermal
equilibrium, $\gamma_{el}$ is the intraspecies collision rate for \K{39}, 
$\bar{n}$ is the average density, $\sigma$ is the scattering cross section,
 $v$ is the relative velocity of two colliding atoms and $\langle \cdot \rangle$ 
indicates averaging over the velocity distribution in the cloud. 

If one indicates the geometric average of the temperature along the 
different direction as $\bar{T}=(T_r^2 T_z)^{1/3}$ and introduces an average,
temperature-dependent cross section defined as:
\begin{equation}\label{eq:sigtilde}
\tilde{\sigma}(\bar{T}) = \frac{\langle \sigma v \rangle}{\langle v \rangle} \qquad ,
\end{equation}
since the average density of a harmonic trap is:
\begin{equation}\label{eq:barn}
\bar{n} = \left( \frac{m}{4 \pi k_B \bar{T}} \right)^{3/2} N_K \omega_r^2 \omega_z
\end{equation}
and the average relative velocity is 
$\langle v \rangle = 4\,(k_B \bar{T})^{1/2}\,(\pi m)^{-1/2}$, the relaxation rate 
can be expressed as:
\begin{equation}\label{eq:gscat}
\tau^{-1} = \frac{\gamma_{el}}{\alpha} = \frac{\tilde{\sigma}(\bar T)}{\alpha} \, N_K \, 
\frac{m \omega_r^2 \omega_z }{2 \pi^2 k_B \bar{T}}.
\end{equation}
We will come back to the relation between $\tilde{\sigma}$ and the actual 
energy-dependent cross-section in Sec. \ref{sec:scatlen}.

Eq.~(\ref{eq:gscat}) holds only if one assumes that the average density of the
sample does not change during the relaxation process. Actually the loss rate 
induced by collisions with atoms of the background gas is about 
$0.03 \, \mathrm s^{-1}$ for the experiments reported in this work. 
If this rate is not negligible as compared to the measured relaxation rates, it must be
taken into account: one expects that the reduction of the density slows down the
relaxation so that after an initial decay with a rate corresponding to
Eq.~(\ref{eq:gscat}), equilibrium apparently sets for a 
value of $E_z / E_r$ lower than $1$.
This picture is further complicated by the unavoidable heating rate
present in the experiment that increases the temperature throughout
the measurements: in our system we measure a different heating rate 
in the axial and radial direction so that it is possible to achieve equilibrium 
also for a value of $E_z / E_r$ greater than $1$.

In principle, one could take these effects into account having the equilibrium 
value of the energy ratio as a free parameter: for our measurements however the 
equilibrium value is almost always consistent with $1$. We therefore choose to 
fix it to $1$ in order to improve the estimation of the time constant, as it is 
shown in Fig. \ref{fig:relax}.

\subsection{Measurements of the elastic cross section}
In order to reduce systematic errors on the determination of $\tilde{\sigma}$, 
we performed several measurements at different densities and for two different 
temperatures of $29$ and $16 \,\mu$K. Furthermore, to check that the 
inhomogeneous magnetic field present in our \emph{in situ} imaging was not a source 
of error, the dataset at $16 \,\mu$K is taken with a slightly different procedure. 
First, we adiabatically decompress the trap to 
$\omega_r = 2 \pi \times 290(1)\,$Hz and $\omega_z = 2 \pi \times 21.24(1)\,$Hz and 
then blow away \Rb, apply parametric heating at $\omega_r$, wait 
for a variable time and image the cloud after a $2\,$ms expansion.

\begin{figure}[!ht]
\centering
\includegraphics[width=220pt]{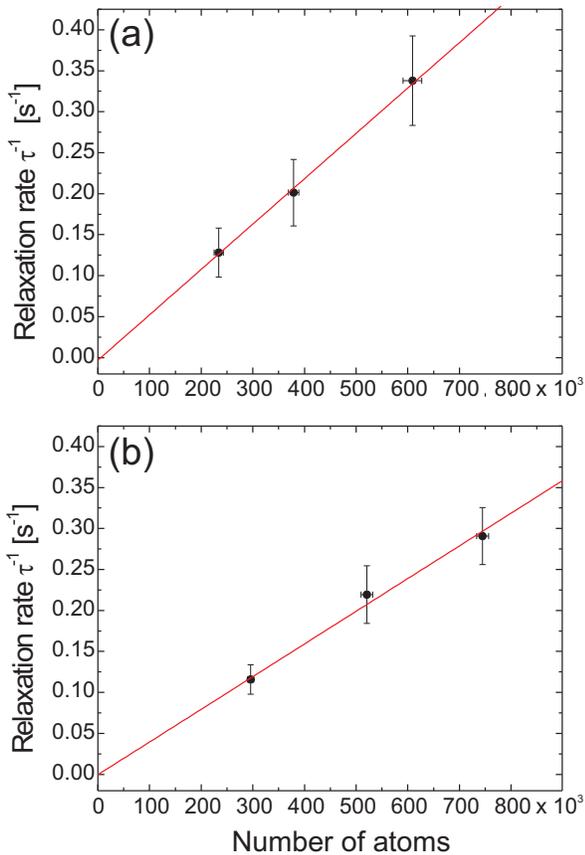}
\caption{(Color online) Plot of the measured relaxation rate as a function of the number 
of atoms in the sample after expansion (a) and \emph{in situ} (b). The measured initial 
average temperature is $16(1)\,\mu$K and $28.9(1.3)\,\mu$K respectively. The slope 
of the linear fit to the data is related to the elastic cross-section, while the 
intercept is a measure of ergodic mixing as described in the text. 
The fit results are: $\gamma_{mix} = -0.003(0.065)\, \mathrm{s}^{-1}$ and 
$A = 5.5(1.6) \times 10^{-7} \, \mathrm{s}^{-1}$ for figure (a) and $\gamma_{mix} = 
-0.0005(0.034)\, \mathrm{s}^{-1}$ and 
$A = 4.0(1.0) \times 10^{-7} \, \mathrm{s}^{-1}$ for figure (b). Error bars on the 
vertical direction are statistical error on the exponential decay fit, while on the
horizontal direction they are the statistical fluctuation of the initial number of
atoms not including the calibration uncertainty.}\label{fig:figona}
\end{figure}

The results of our measurements are shown in Fig. \ref{fig:figona} where we 
report the relaxation rate $\gamma = \tau^{-1}$ as a function of the
initial number of atoms $N_K$. Figure \ref{fig:figona}(a) shows the 
measurements taken after expansion with an average initial temperature 
$\bar{T} = 16(1)\,\mu$K, while Fig. 
\ref{fig:figona}(b) presents measurements taken \emph{in situ} with 
$\bar{T} = 28.9(1.3)\,\mu$K. One can see that the relaxation rate has the expected linear
dependence on the number of atoms and the extrapolation towards zero is consistent with 
zero.

In a Ioffe-Pritchard trap, a sufficiently big atomic cloud can experience a potential
that is not strictly separable so that different degrees of freedom are coupled and 
relaxation can occur even in the absence of collisions. We refer to this process as 
\emph{ergodic mixing}. \emph{A priori}, for equal harmonic frequencies, ergodic mixing
could play a more prominent role in our mTrap than in the usual traps due to 
its small size. For this reason we take ergodic mixing into account by separating the 
component of the relaxation rate, which is linear in $N$, from the extrapolation in the 
limit of zero density where relaxation can only occur through ergodic mixing. One can 
therefore assume that
\begin{equation}\label{eq:line}
\gamma = \alpha^{-1} \frac{\mathrm d \gamma_{el}}{\mathrm d N_K} N_K + \gamma_{mix} = 
A N_K + \gamma_{mix} \qquad , 
\end{equation}
where $\gamma_{mix}$ is the time constant of the ergodic mixing process. The two 
parameters $A$ and $\gamma_{mix}$ are obtained from a linear fit of the experimental 
data, as shown in Fig. \ref{fig:figona}. 
By using Eq.~(\ref{eq:gscat}) one has
\begin{equation}\label{eq:A}
A = \frac{\tilde{\sigma}}{\alpha} \, \frac{m \omega_r^2 \omega_z }{2 \pi^2 k_B \bar{T}} \qquad .
\end{equation}
As pointed out above, since the values for $\gamma_{mix}$ are consistent with
zero in both the datasets, we can conclude that ergodic mixing plays a negligible role in
our measurements.

The resulting values for the temperature-dependent average elastic cross section, 
computed from Eq.~(\ref{eq:A}) with $\alpha = 2.7$, as obtained from numerical simulation 
described in Sec. \ref{sec:sim}, are $2.2(0.8) \times 10^{-12}\,\mathrm{cm}^2$ and 
$0.91(0.22) \times 10^{-12}\,\mathrm{cm}^2$ for $16$ and $29 \, \mu$K respectively. These 
values are a very good approximation of the actual cross section as we show in the next 
section.

\subsection{Extracting the value of the scattering length}\label{sec:scatlen}
As pointed out above, the measured cross sections depend on the temperature. 
In order to extract information on the interatomic potential, one has to make some
assumption on the behavior of the cross section as a function of temperature.
A simple partial wave expansion of the scattering amplitude shows that, for the 
experimental condition reached in this work, collisions can only occur at $l=0$ 
angular momentum, odd values being suppressed on symmetry ground and even values 
due to the low temperature \footnote{At $30 \,\mu$K the $d-$wave contribution is $5$
order of magnitude smaller (A. Simoni, private communication (2006)).}.
In a very general form the s-wave cross section for identical 
boson is:
\begin{eqnarray}
  \sigma(k) &=& 8 \pi |f_0(k)|^2
\end{eqnarray}
where $k = m v / (2 \hbar)$ is the relative wavevector and $f_0(k)$ is the s-wave scattering 
amplitude that can be expressed in terms of the S-matrix phase shift 
$\delta_0(k)$: $f_0(k)=e^{i \delta_0(k)}\sin \delta_0(k) /k$.

In the limit of vanishing energy, given that the potential decays more rapidly
than $1/r^3$, we can express the cross section as a function of a single parameter,
the s-wave scattering length which, for two alkali atom in the stretched 
$\ket{F = 2, m_F=+2}$ state, is the triplet one $a_T$ :
\begin{eqnarray}
|f_0(k)|^2& \simeq & a^2_T\\
\sigma(k)& \simeq& 8 \pi a^2_T
\end{eqnarray}

For higher temperature one has to calculate the scattering amplitude up to order 
$k^2: f_0(k) = (-1/a_T +i k +\frac{1}{2}k^2 r_e + \dots )^{-1}$, 
where $r_e$ denotes the effective range, which is a function of the scattering 
length $a_T$ and the $C_6$ coefficient of the van der Waals potential 
\cite{FlambaumEffRange}
\begin{eqnarray} \label{sigmavsat}
\sigma(k)= 8 \pi \frac{a^2_T}{(1-\frac{1}{2}k^2 r_e a_T)^2+k^2 a_T^2}.
\end{eqnarray}

Since the $C_6$ coefficients are well known for all alkali dimers 
\cite{DereviankoC6,BohnC6}, 
we numerically invert Eq.~(\ref{eq:sigtilde}) after inserting Eq.~(\ref{sigmavsat}),
averaging over the Boltzmann thermal distribution and assuming $a_T < 0$, to 
obtain the scattering length: 
$a_T = -67(11)a_0$ from data at $16 \,\mu$K and $a_T = -48(5) a_0$ from data at  
$29 \,\mu$K \footnote{We note that, at this temperature, $\tilde{\sigma}$ differs
from the thermal average of Eq.~(\ref{sigmavsat}) by 14\%.}. For comparison, we notice that, if we neglect the effective
range, these values would be $-57(11)a_0$ and $-36(5) a_0$, respectively. 
In the first case the two values differ by approximately 1.57 combined standard
deviations, while in the second case the difference is 1.75.
The quoted uncertainties derive mainly from the fits of Fig. \ref{fig:figona} 
and the atom number calibration ($\pm 20\%$), which was done independently for each data
set. Finally, we take a weighted average of the two scattering length values calculated
with effective range and multiply the associated uncertainty by a factor $\sqrt{\chi^2}=1.65$, to set 
the confidence level to 68\% \cite{BrandtData}: the result is $a_T = -51(7)a_0$.

Prior to this work, the \K{39}\ triplet s-wave scattering length has been measured 
by photoassociation spectroscopy, $a_T= -33(5) a_0$ \cite{Gould39K}, and by mass-scaling 
from rethermalization of \K{40}, $a_T= -37(6) a_0$ \cite{JinFeshbachK40}. Indeed our
measurement is the first direct determination  of the elastic cross section for this 
isotope. To different extents, all reported values of the scattering length depend on 
the $C_6$ coefficient of the van der Waals long range potential: we take $C_6=3927\,$a.u.
\cite{BohnC6}, while in Ref.~\cite{Gould39K,JinFeshbachK40} $C_6$ is assumed
equal to $3897\,$a.u. \cite{DereviankoC6}. 
We checked that our value changes with $C_6$ with a rate 
$\delta |a_t|=0.001 a_0 \times \delta C_6$ [a.u.], approximatively $50$ times smaller 
than in Ref.~\cite{Gould39K} (no such rate is available in Ref.~\cite{JinFeshbachK40}) 
and therefore the difference in the $C_6$ does not significantly change 
the derived scattering length. As our result differs from the average of the published 
values of $1.70$ combined standard deviation, we conclude that the agreement is 
satisfactory.

\subsection{Numerical simulations}\label{sec:sim}
As shown in Eq.~(\ref{eq:A}), to compute the value of the scattering 
length from the fit to the data one needs to know the value of the parameter
$\alpha$, which is the average number of collisions per particle needed for
thermalization. This parameter depends on the temperature and the trap 
frequencies in a non trivial way and can vary from 2.4 to 3.4 \cite{DalibardVarenna}.
Following \cite{MonroeCsColl, DalibardCsCollision, FootDMCS} we estimate this 
parameter from a numerical simulation of the system. Taking advantage of the low 
density and small atom number of the sample, we make a direct simulation of the gas 
in which we consider 3D position and velocity of every single atom. Choosing
a time step $\delta t$ much smaller than both the average collision time 
$\gamma_{coll}^{-1}$ and the faster timescale of single particle dynamics, namely 
$\omega_r^{-1}$, one may assume that during the time interval $\delta t$ 
interactions between atoms are weak enough that they decouple with the center 
of mass motion and that the force experienced by the atoms varies little. 
Therefore the dynamics of the gas can be directly simulated with the following procedure:
\begin{enumerate}
\item Position and velocity are updated according to external forces with a Verlet 
integrator.
\item Every $M$ steps, the positions of the atoms in real space are discretized on a
lattice of spacing $\delta x$ and if two atoms are located at the same site a collision
test is made.
\item If the collision test is positive, the collision is resolved using simple 
classical mechanics and s-wave scattering: 
$$ \left\{ \begin{array}{c}
\mathbf v_1' = \mathbf v_{CM} + \frac{|\mathbf v_r|}{2} \, \hat{\mathbf e}_R \\
\mathbf v_2' = \mathbf v_{CM} - \frac{|\mathbf v_r|}{2} \, \hat{\mathbf e}_R 
\end{array} \right. $$ 
where $\mathbf v_i'$ indicates the velocity of particle $i$ after the collision, 
$\mathbf v_{CM}$ and $\mathbf v_r$ are the center of mass and relative velocity 
respectively, before the collision, and $\hat{\mathbf e}_R$ is a random
direction on the unit sphere.
\end{enumerate} 
The collision test consists in comparing a random real number $y$, uniformly distributed
in $\left[0, 1\right)$, with the collision probability calculated according to kinetic
theory:
\begin{equation}
\wp = \sigma \, |v_r| \, M\,\delta t \, \delta x^{-3}
\end{equation}
where $\sigma$ is the elastic scattering  cross section and $|v_r|$ is the modulus of the 
relative velocity; collision is processed if $y < \wp$. In order to reproduce a correct
scattering rate, one must choose $\delta x$ so that the average number of atoms in a 
volume $\delta x^3$ is much smaller than $1$ and $M$ so that the distance traveled by an 
atom during the time $M\,\delta t$ is of the order of $\delta x$.

\begin{figure}
\centering
\includegraphics[width=220pt]{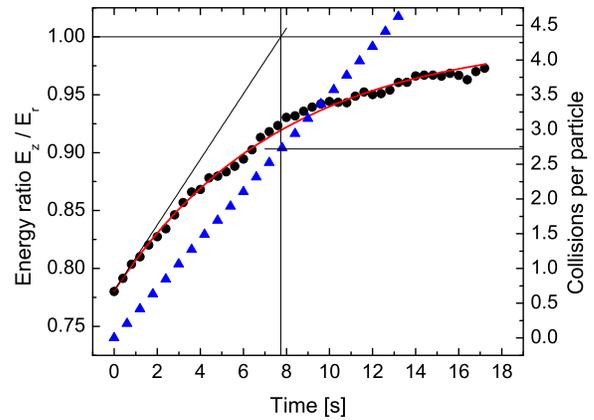}
\caption{(Color online) Plot of the simulated energy ratio (black dots, left scale) 
and collisions per particle (blue triangles, right scale) as a function of time for 
$20 \times 10^3$ atoms in a harmonic trap whose frequencies are $440$ and $29.4\,$Hz
in the radial and axial direction respectively. The red curve is an exponential
decay fit. The value $\alpha = 2.72$ can be obtained directly from the
graph identifying the time constant of the decay ($7.74$ s).}\label{fig:sim}
\end{figure}
In Fig.~\ref{fig:sim} we report a typical result of the simulation showing the radial 
to axial energy ratio and the number of collisions per atom as a function of time. The
simulation is performed in a cylindrical harmonic trap with radial and axial frequency of
$440$ and $29.4\,$Hz respectively. The number of atoms is $2\times 10^4$ and the initial 
energy distribution has a width of $15(19.5)\, \mu$K along the axial (radial) direction.
The cross section is $8.6 \times 10^{-16}\,\mathrm{m}^2$ so that the scattering rate in
the simulation is close to the experimental one. The value of $\alpha = 2.7$ can be
extracted directly from the graph as explained in the caption of Fig.~\ref{fig:sim}.

\section{Conclusions}
In summary we have demonstrated sympathetic cooling of \K{39}\ with \Rb, 
in spite of the low interspecies cross section, and we have measured the
cross section for elastic collision between two \K{39}\ atoms in a pure 
triplet state. We have hence obtained the triplet s-wave
scattering length $a_T$, assuming its negative sign and an effective 
range approximation for the s-wave scattering amplitude at low but 
finite temperature.

The possibility of sympathetic cooling and the knowledge of the 
collisional properties of bosonic potassium open the way to its 
use as a quantum degenerate gas. Bosonic potassium isotopes, 
\K{39}\ and \K{41}, are predicted to feature 
Feshbach resonances, several Gauss wide, at moderate fields ($< 1$~kG). 
As a consequence, optically trapped bosonic potassium appears a suitable 
candidate to realize a condensate with a scattering length tunable around
zero. Such a condensate would be interesting for interferometric purposes,
as the interaction energy limits the accuracy of interferometers 
employing Bose-Einstein condensates. In addition, wide Feshbach resonances 
make potassium condensates particularly attractive to realize double
condensates in optical lattices, for quantum simulation purposes.
For \K{39}, in particular, Bohn and coworkers have predicted a Feshbach 
resonance around $40$ G for two atoms in the $\ket{F=1,m_F=-1}$ state, with 
a width of several tens of Gauss \cite{BohnKScat}, while another 60 G 
broad resonance has been predicted around 400 G for atoms in the absolute 
ground state $\ket{F=1,m_F=1}$ \cite{tiesinga}. While a double condensate of 
\K{41}\ and \Rb\ was produced in a magnetic trap \cite{ModugnoTBEC}, \K{39}\ 
 has not been brought to degeneracy so far. This work represents a decisive 
step in this direction as we reached a phase-space density of $0.01$, with a gain 
of several orders of magnitude with respect to the state-of-the-art \cite{FortBambini}.

\begin{acknowledgments}
We thank all the degenerate gas group at LENS for many stimulating discussions and
fruitful advice and A. Simoni for useful data on interatomic potential.
This work was supported by MIUR, by EU under Contracts HPRICT1999-00111 and 
MEIF-CT-2004-009939, by INFN through the project SQUAT and by Ente CR Firenze.
\end{acknowledgments}


\bibliographystyle{apsrev} 
\bibliography{lds_bib}  

\end{document}